\documentclass[prl,aps,twocolumn,showpacs,amsmath,amssymb,floatfix]{revtex4}
\pdfoutput=1
\usepackage{graphicx}% Include figure files
\usepackage{dcolumn}% Align table columns on decimal point
\usepackage{bm}% bold math
\usepackage{textcomp}
\usepackage{amsmath}

\begin{document}
 \noindent

\title{Quantum jumps and spin dynamics of interacting atoms in a strongly coupled atom-cavity system}
\author{M.~Khudaverdyan}\email{mika@iap.uni-bonn.de}
    \author{W.~Alt}
    \author{T.~Kampschulte}
    \author{S.~Reick}
    \author{A.~Thobe}
    \author{A.~Widera}
    \author{D.~Meschede}

 \affiliation{Institut f\"{u}r Angewandte Physik, Universit\"{a}t Bonn, Wegelerstr. 8, 53115 Bonn, Germany}

\begin{abstract}
We experimentally investigate the spin dynamics of one and two
neutral atoms strongly coupled to a high finesse optical cavity.
We observe quantum jumps between hyperfine ground states of a
single atom. The interaction-induced normal mode splitting of the
atom-cavity system is measured via the atomic excitation.
Moreover, we observe the mutual influence of two
atoms simultaneously coupled to the cavity mode. 
\end{abstract}

\maketitle

The strong coherent light-matter interaction induced by a high
finesse resonator allows to study and manipulate neutral atoms at
the quantum level \cite{HarocheBook}. Information about the
evolution of the coupled system is contained in both the resonator light
field and the atom. In particular, the dynamics of a single atom
coupled to the cavity can be inferred with a high information rate
from the light field. Continuous monitoring of the atomic motion
and the atom number have already been demonstrated
\cite{Hood00,Pinkse00,McKeever04}. However, the dynamics of the
internal spin states, which plays a central role in many proposals
to engineer entangled quantum states
\cite{CorrelationsProb,CorrelationsDeter}, has so far not been
resolved.

Here we observe the internal dynamics of one and two atoms in a
high finesse optical resonator. Continuous observation of a
quantum system can reveal quantum jumps, i.e.~the sudden and
random change of a quantum system's state over time due to
interaction with the environment, as has been observed in various
systems \cite{Ionjumps,basche1995,Peil1999,Gleyzes2007}. Here we
use the cavity to directly observe the spin quantum jumps of a
single atom, as anticipated in Ref.~\cite{McKeever04}. This method
is based on the suppression of the cavity transmission
\cite{Boozer06} due to strong coherent interaction, leading to
normal mode splitting. For single atoms this splitting has been
observed by measuring the intra-cavity photon number
\cite{Maunz2005,Boca2004}. In contrast, we use the cavity-based
detection of the atomic state to measure the normal mode splitting
via the atomic part of the excitation. Extending our experiment to
the simultaneous coupling of two atoms to the cavity mode we
observe the coupled dynamics of the atomic states, as the state of
each atom influences the intra-cavity photon number experienced by
the other atom.

%%%%%%%%%%%%%%%%%%%%%%%%%%%%%%%%%%%%%%%%%%%%%%%%%%%%%%%%%%%%%%
%%%%%%%%%%%%%%%          Figure 1         %%%%%%%%%%%%%%%%%%%%
%%%%%%%%%%%%%%%%%%%%%%%%%%%%%%%%%%%%%%%%%%%%%%%%%%%%%%%%%%%%%%
\begin{figure}[!t]
        \centering
        \includegraphics[width=0.98\columnwidth]{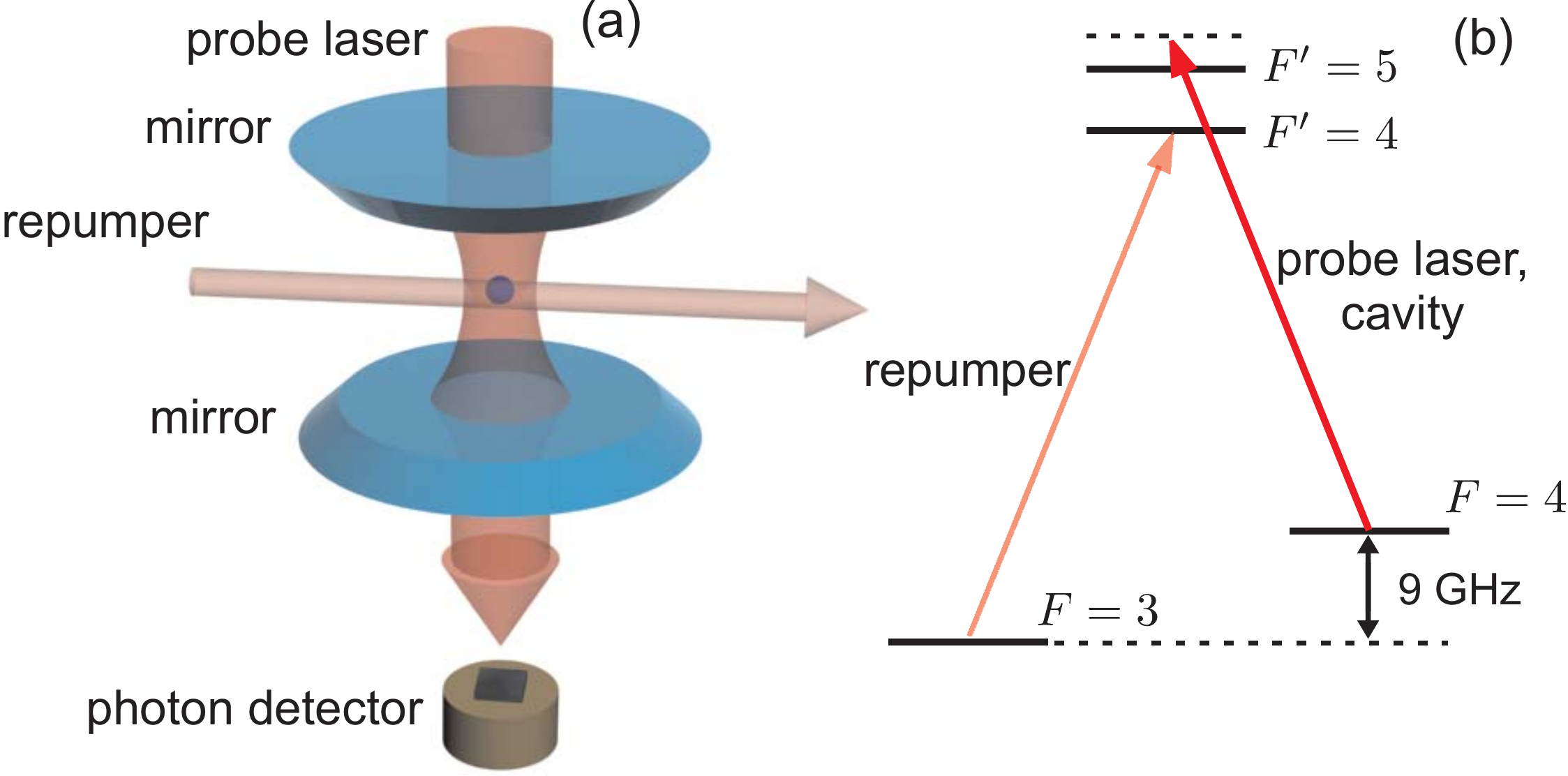}
\caption{(a) A single atom is placed into the cavity mode. The
state of the atom is continuously monitored by observing the
transmission of the probe laser, tuned close to the \mbox{$F=4
\longrightarrow F'=5$} transition. A similar scheme applies to two
atoms. (b) Level scheme of relevant energy levels and transitions
for Cs.}
        \label{fig:concept}
\end{figure}
%%%%%%%%%%%%%%%%%%%%%%%%%%%%%%%%%%%%%%%%%%%%%%%%%%%%%%%%%%%%%%
%

In our approach neutral Caesium (Cs) atoms are coupled to the mode
of an optical resonator as depicted in Fig.~\ref{fig:concept}. We
monitor the transmission of a probe laser beam with angular
frequency $\omega_\mathrm{p}$, tuned to the resonance frequency of
the cavity $\omega_\mathrm{c}=\omega_\mathrm{p}$. The cavity
itself is blue-detuned by
$\omega_\mathrm{c}-\omega_\mathrm{4,5'}=2 \pi\times44$~MHz from
the \mbox{$F=4 \longrightarrow F'=5$} transition of the Cs $D_2$ line
at 852\,nm \cite{footnote}, see Fig.~\ref{fig:concept}. Here
$\omega_\mathrm{4,5'}$ is the angular frequency of the
\mbox{$F=4\longrightarrow F'=5$} transition, where $F$ is the
total atomic angular momentum. The cavity transmission level is a
direct and continuous measure of the atomic state: An atom in the
$F=3$ hyperfine state does not couple to the resonator mode, hence
the laser beam is fully transmitted. An atom in the $F=4$ state,
however, couples strongly to the cavity mode and leads to a
normal-mode splitting of the cavity resonance. This effectively
blocks the transmission of the probe laser beam.

%%%%%%%%%%%%%%%%%%%%%%%%%%%%%%%%%%%%%%%%%%%%%%%%%%%%%%%%%%%%%%
%%%%%%%%%%%%%%%%%%%%%%%%%%%%%%%%%%%%%%%%%%%%%%%%%%%%%%%%%%%%%%
%\paragraph*{Experimental details ---}
At the beginning of each experimental sequence we load a single
laser cooled Cs atom into a standing-wave optical dipole trap at
1030\, nm. Using the trap as an optical conveyor belt we transport
the atom to a well defined position within the field of the
cavity, for details see \cite{Khudaverdyan2008}. The cavity mode
has a diameter of $2\omega_0=46\,\mu$m and a length of $159~\mu$m.
The parameters of our atom-cavity system are $\{g,\kappa,\gamma\}=
2 \pi \times \{8 \ldots 13, 0.4, 2.6\} \,$MHz. The expected
atom-cavity coupling strength $g$ varies for different Zeeman
sublevels of the $F=4 \longrightarrow F'=5$ transition, $2\gamma$ and $2\kappa$
are the linewidths of the $D_2$ transition and cavity resonance, respectively. 
Thus our experiment operates in the strong coupling regime with a single atom
cooperativity parameter $C_1=g^2/(2\kappa\gamma)>30$.

\begin{figure}[!ct]
        \centering
        \includegraphics[width=0.98\columnwidth]{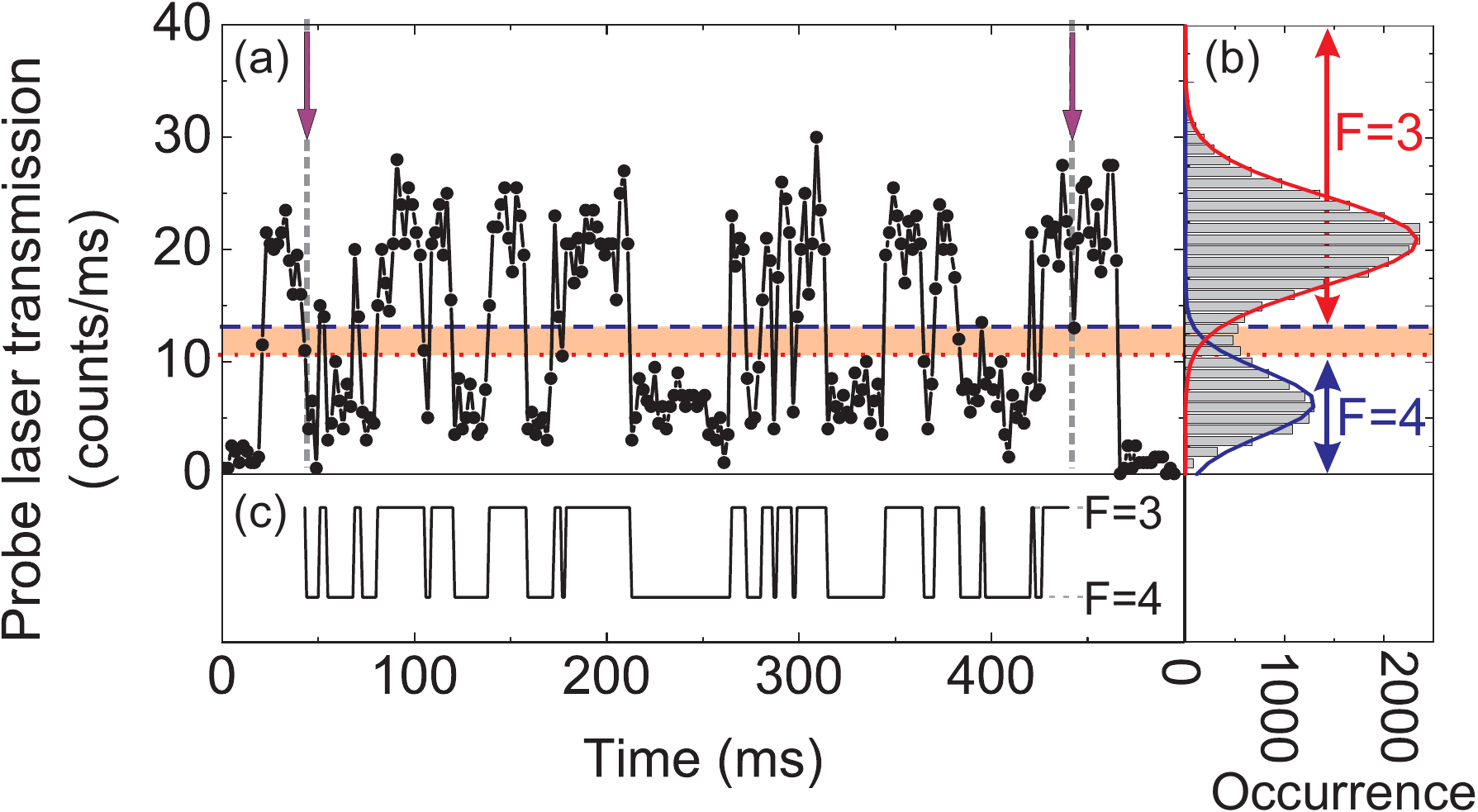}
\caption{(a) Measured cavity transmission over time, exhibiting
quantum jumps of a single atom between $F=3$ and \mbox{$F=4$}.
Arrows indicate the insertion and removal of the atom. (b)
Histogram of the transmission levels. The two peaks represent the
$F=3$ and \mbox{$F=4$} states, the horizontal dashed (dotted) line
marks the thresholds for the spin $F=3$ ($F=4$) state (see text).
(c) Reconstructed atomic state for the transmission signal in (a).
}
        \label{fig:state detection}
    \end{figure}

%%%%%%%%%%%%%%%%%%%%%%%%%%%%%%%%%%%%%%%%%%%%%%%%%%%%%%%%%%%%%%
%\paragraph*{Observation of quantum jumps ---}
When we observe the cavity transmission continuously, a random
telegraph signal originates from quantum jumps of a single atom
between the $F=4$ (low transmission signal) and $F=3$ (high
transmission signal) hyperfine states over about 400\,ms (see
Fig.~\ref{fig:state detection}a). The intra-cavity photon
number for high transmission is equal to 0.3. With an
overall photon detection efficiency of 1.3\% the resulting count
rate at the photon detector is about 20\,counts/ms. The
transitions from the $F=4$ to the $F=3$ state are caused by the
cavity field, off-resonantly exciting the atom to the $F'=4$
state, from which it decays into the $F=3$ ground state. For atoms
in the $F=3$ state the rate of quantum jumps is strongly reduced
due to the large detuning of the probe laser from the
\mbox{$F=3\longrightarrow F'=4$} transition. To induce transitions
from the $F=3$ to the $F=4$ state at a comparable rate, we apply
an additional repumping laser which is resonant with the
\mbox{$F=3\longrightarrow F'=4$} transition (see
Fig.~\ref{fig:concept}). 
Although strong repumping lasers have been applied in previous work with single atoms \cite{Pinkse00,Boozer06}, for our repump intensity quantum jumps can be resolved with our time resolution of 2\,ms, given by the binning time.

%%%%%%%%%%%%%%%%%%%%%%%%%%%%%%%%%%%%%%%%%%%%%%%%%%%%%%%%%%%%%%
%%%%%%%%%%%%%%%          Figure 2         %%%%%%%%%%%%%%%%%%%%
%%%%%%%%%%%%%%%%%%%%%%%%%%%%%%%%%%%%%%%%%%%%%%%%%%%%%%%%%%%%%%
%%%%%%%%%%%%%%%%%%%%%%%%%%%%%%%%%%%%%%%%%%%%%%%%%%%%%%%%%%%%%%
%\paragraph*{State detection details ---}
In order to extract the atomic state from the transmission signal
we form a histogram from 163 telegraph traces (see
Fig.~\ref{fig:state detection}(b)) which shows two peaks,
reflecting the $F=3$ and $F=4$ spin states, respectively. We fit
each peak with a Gaussian distribution. The two peaks partially
overlap, so that in this range the spin state cannot be determined
unambiguously. We therefore define the region of
transmission values corresponding to the $F=4$ state, denoted in
Fig.~\ref{fig:state detection} (b) by the blue arrow, such that
only 1\% of the values for an atom in $F=3$ lies in this region.
Vice versa, the region labeled by the red arrow is chosen to
contain only 1\% of the transmission values related to the $F=4$
state and is therefore used to indicate $F=3$. In the overlap
region, which contains about 4\% of all time bins, the ambiguity
is resolved if the previous and subsequent transmission values
correspond to the same spin state. In this case the middle value
is considered as noise, as it would otherwise correspond to two
directly subsequent quantum jumps which are much less probable. If
previous and subsequent values are different, the middle value in
question is interpreted as a quantum jump. This allows us to
reconstruct the atomic state, see Fig.~\ref{fig:state detection}
(c).

%\paragraph{Analysis of the dynamics ---}
We characterize the dynamics of the quantum jumps by analyzing the
second order correlation function of the reconstructed atomic
state as proposed in Ref.~\cite{cook1985}, and thereby determine
the rates of the quantum jumps. From the correlation function we
find the rate for a quantum jump from $F=4$ to $F=3$ to be about
106\,s$^{-1}$, set by the intensity of the probe laser field
inside the cavity. Likewise, the rate of jumps from $F=3$ to $F=4$
is found to be 42\,s$^{-1}$, corresponding to the applied
repumping laser intensity.

We use this atomic spin detection method to deduce the normal mode
splitting from the decay of the \emph{atomic} excitation, rather
than the \emph{photonic} excitation.
%%%%%%%%%%%%%%%%%%%%%%%%%%%%%%%%%%%%%%%%%%%%%%%%%%%%%%%%%%%%%%
%\paragraph*{Normal mode splitting sequence ---}
In the first step of the measurement, we induce the normal mode
splitting by coupling an atom in the $F=4, m_{F}=4$ state to the
resonator, this time tuned close to the \mbox{$F=4 \longrightarrow
F'=4$} transition. Here, $m_{F}=4$ is the projection of the
angular momentum onto the quantization axis. The atom-cavity
system is then probed with a weak $70\,\mu$s probe laser pulse
with variable detuning $\omega_\mathrm{p}-\omega_\mathrm{c}$, adjusted to not saturate the population transfer to $F=3$. If
the atom gets excited to the $F'=4$ state, it can decay to the
$F=3$ and $F=4$ ground states with comparable probabilities.
Consequently, the atomic excitation probability is mapped onto the
population of the dark $F=3$ ground state. Remarkably, during this
process the atom scatters on average only two photons, therefore
experiencing negligible heating as it decouples from the cavity in
the $F=3$ state. In a second step, we shift the cavity resonance
40\,MHz to the blue of the \mbox{$F=4 \longrightarrow F'=5$}
transition and record the transmission of the resonant probe
laser. From the first 2\,ms of the transmission level we deduce
the hyperfine state of the atom as described above. The typical
duration of one experimental cycle is 35\,ms, and it is repeated
forty times for one atom. This leads to a high information rate
exceeding typical rates when the atomic state is measured with
push-out techniques \cite{Pushout}. Due to
thermal oscillations and varying position, the atom experiences
different coupling strengths. Recording the transmission level for
10\,ms with the repumping laser switched on at the end of each
measurement cycle, we post select only those events with a strong
atom-cavity coupling, i.e.~where the transmission level lies below
30\,\% of the empty cavity transmission before and after probing
the normal mode splitting.

\begin{figure}[!t]
        \centering
        \includegraphics[width=0.98\columnwidth]{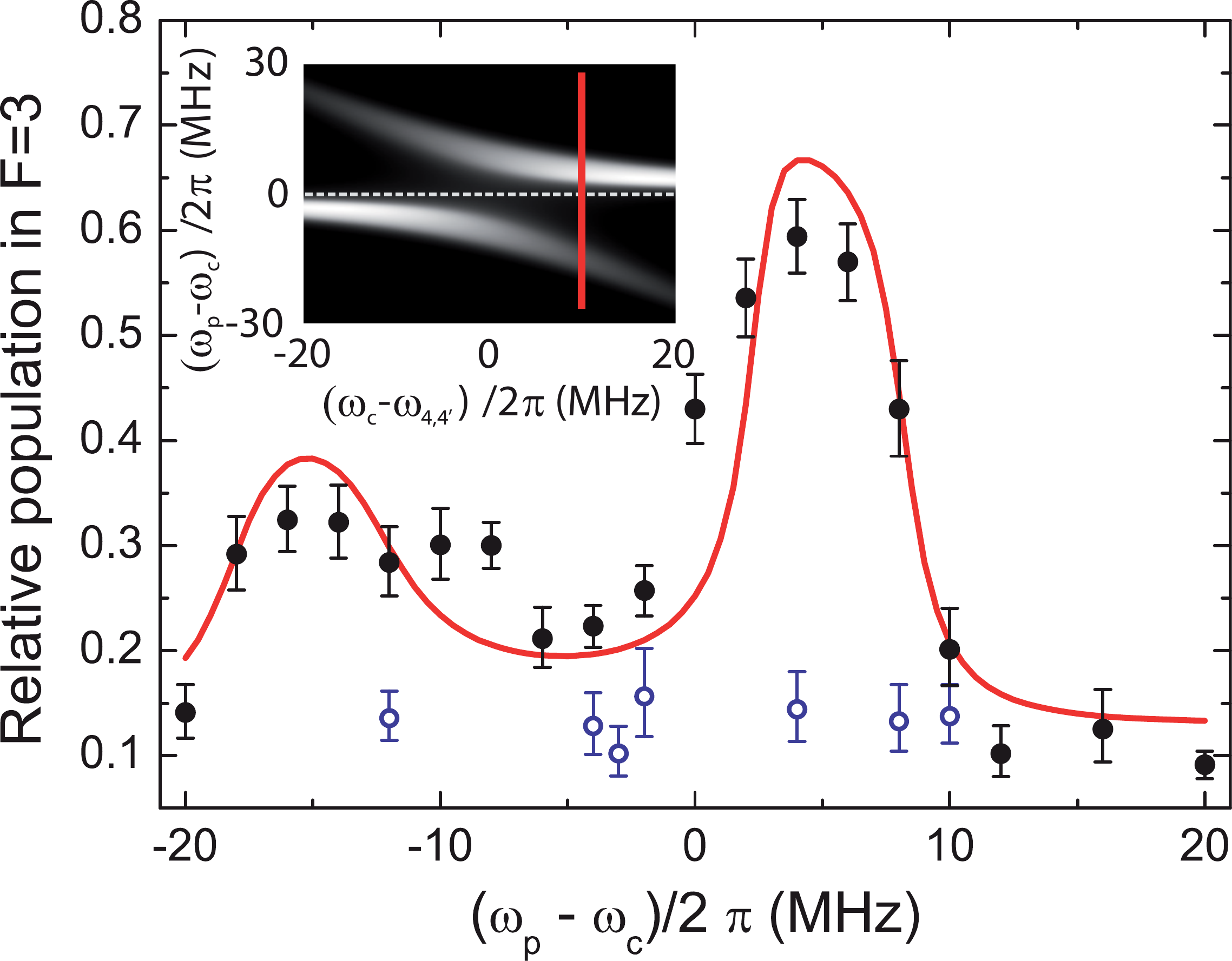}
\caption{Normal mode splitting, measured via population in
$F=3$ versus probe-cavity detuning. Each data point ({\large
$\bullet$}) is the result of 120 to 470 experimental cycles.
Erroneous detection of an atom in $F=3$ causes a
background of 13\% ({\Large $\circ$}). Each such point is measured immediately after the data point for a certain detuning and therefore plotted at this detuning value. The solid curve is
the result of our model (see text). The inset shows the calculated
scattering rate of a single atom inside the cavity.}
        \label{fig:VRS}
    \end{figure}

The population in \mbox{$F=3$} as a function of the detuning
$\omega_\mathrm{p}-\omega_\mathrm{c}$ is shown in
Fig.~\ref{fig:VRS}, revealing the interaction-induced normal-mode
splitting. Here, the cavity-atom detuning is $\omega_\mathrm{c} -
\omega_\mathrm{4,4'} = 2 \pi\times 10\,$MHz, extracted from the
numerical model (see below), where $\omega_\mathrm{4,4'}$ is the
frequency of the \mbox{$F=4 \longrightarrow F'=4$} transition. By performing the same sequence without the $70\,\mu$s mapping pulse we detect a
background of approximately 13\% erroneous
detection events of the atomic state during the 2\,ms state detection
time. This value is compatible with the probability of 18\% of a
quantum jump to occur during the detection interval.

We analyze our measurement with a simple model: First, we estimate
the photon scattering rate of the atom as a function of the
cavity-probe laser detuning by numerically solving the master
equation \cite{Carmichael}. With this scattering rate we
employ a rate equation to model the nonlinear population transfer
to the $F=3$ state during the 70 $\mu$s mapping pulse. For this
we assume a three-level system comprised of the $F=3$
and $F=4$ ground states and the $F^\prime = 4$ excited state,
neglecting population redistribution over Zeeman sublevels. We
assume a homogeneous distribution of $g=2 \pi \times (6 \ldots
12)$\,MHz, which corresponds to the selected range of
transmissions mentioned above. Here, $g = 2\pi \times12\,$MHz
belongs to a maximally coupled atom for the \mbox{$F=4, m_{F}=4
\longrightarrow F'=4, m_{F'}=4$} transition. 
Fitting the model to the measured data yields as two fit parameters the
empty cavity photon number $n_\mathrm{ph}= 0.062(3)$ and the cavity-atom detuning $\omega_\mathrm{c} -
\omega_\mathrm{4,4'}=2 \pi \times 10(1)\,$MHz (solid line in inset of
Fig.~\ref{fig:VRS}). An independent photon number measurement
agrees with the fitted value and implies that we are in the weak
excitation limit.

%
%%%%%%%%%%%%%%%%%%%%%%%%%%%%%%%%%%%%%%%%%%%%%%%%%%%%%%%%%%%%%%
%%%%%%%%%%%%%%%          Figure 6         %%%%%%%%%%%%%%%%%%%%
%%%%%%%%%%%%%%%%%%%%%%%%%%%%%%%%%%%%%%%%%%%%%%%%%%%%%%%%%%%%%%
%%%%%%%%%%%%%%%%%%%%%%%%%%%%%%%%%%%%%%%%%%%%%%%%%%%%%%%%%%%%%%
%
%\paragraph*{Atom number dependency ---}
Extending our experiment to the case of two atoms coupled to the
resonator, each atom affects the light field in the resonator
which is experienced by the other atom. 
The probability of a quantum jump to occur from $F=4$ to $F=3$ then depends
on the number of atoms in $F=4$. Thus, the dynamics of quantum
jumps becomes conditional on the state of both atoms.
While in our current experiment the coherent evolution of a quantum state due to this interaction cannot be observed due to photon scattering, the effective interaction does change the atomic spin dynamics.

\begin{figure}[!t]
        \centering
        \includegraphics[width=0.98\columnwidth]{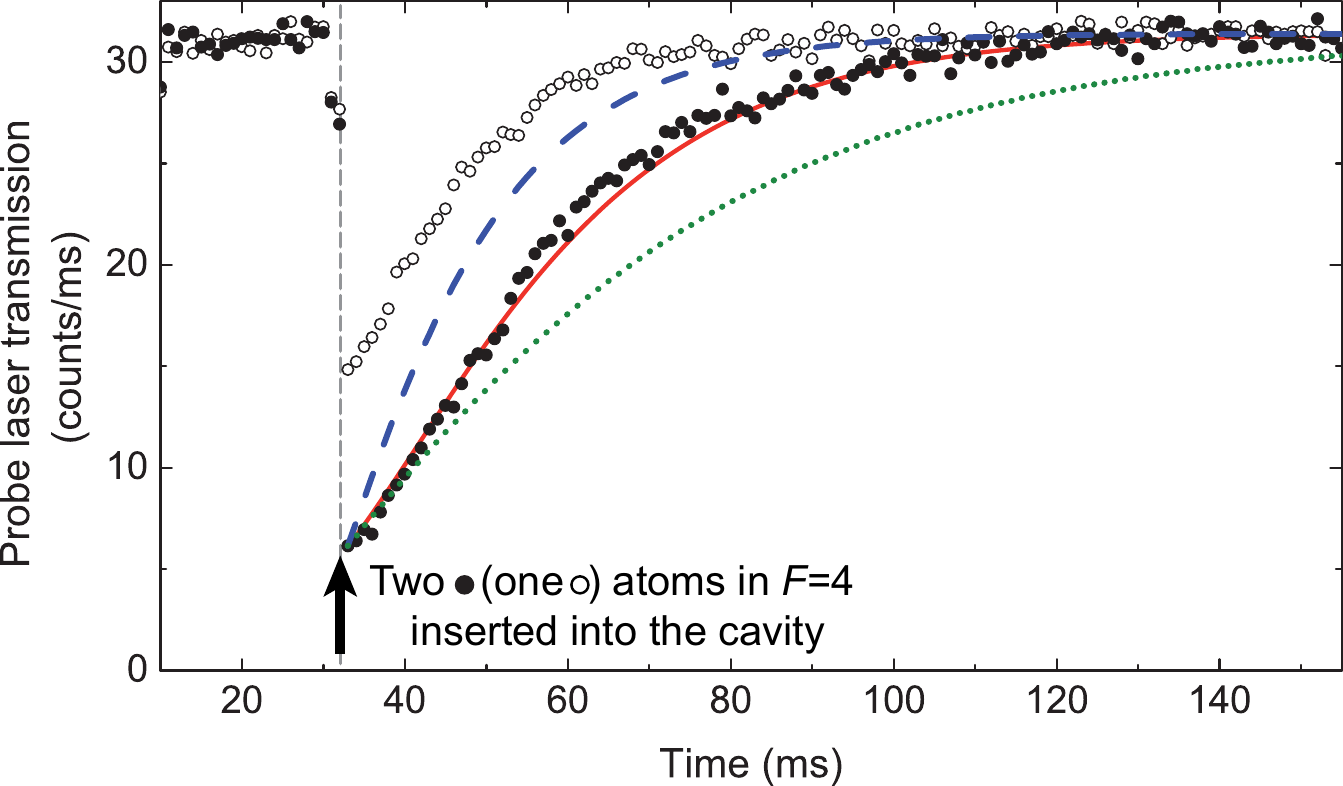}
\caption{Averaged cavity transmission for one ({\large $\circ$})
and two atoms ({\large $\bullet$}) simultaneously coupled to the
cavity. The solid line is the theoretical expectation for coupled dynamics of two atoms. For
comparison, the dashed (dotted) line assumes an atom number
independent rate of quantum jumps $R=R_{1}$
($R=R_{2}$). The arrow indicates the insertion of atoms. The data has been taken with improved detection efficiency of 4.5\% and a binning time of 1\,ms.}
        \label{fig:QJTwoAtoms}
    \end{figure}

We check that the distance between two atoms is below
2\,$\mu$m, smaller than the waist of the cavity. We then position
both atoms about $15\,\mu$m away from the cavity axis, and pump them to
$F=4$. There, for our parameters, the intra-cavity photon number
depends sensitively on the number of atoms in $F=4$
\cite{Khudaverdyan2008,footnote1}. We monitor the probe laser
transmission for about 120\,ms while the repumping laser is
switched off. Averaging over 169 of such traces we obtain the
ensemble average shown in Fig.~\ref{fig:QJTwoAtoms}. This dynamics
of the transmission signal is well explained without free
parameters by assuming that the rate of quantum jumps depends on
the state of both atoms. For a single atom in $F=4$, the rate of
quantum jumps is measured independently to be
 $R_{1}=68\pm2\,\mathrm{s}^{-1}$(see Fig.~\ref{fig:QJTwoAtoms}). When both
atoms are in $F=4$, the quantum jump rate for each atom $R_2$ is
extracted by comparing the initial transmission levels for one and
two coupled atoms, yielding  $R_{2}=28 \pm 1\,\mathrm{s}^{-1}$. Taking this
into account, we obtain the theoretical transmission dependence
depicted in Fig.~\ref{fig:QJTwoAtoms}. It
agrees well with the measured data and confirms the coupled
dynamics of atomic spins, while the assumption of an atom number
independent rate yields an inconsistent behavior. Our signal is an evidence for an effective atom-atom interaction, where the measurement method relies on photon scattering thus objecting the creation of entangled states \cite{CorrelationsProb}. The non-linear dependence of the intra-cavity photon number on the number of atoms in $F=4$, however, is the basis for a conditional phase shift. The resulting dynamics can deterministically induce   entanglement of the two atoms \cite{Lin08}.

%%%%%%%%%%%%%%%          Figure 7         %%%%%%%%%%%%%%%%%%%%
%%%%%%%%%%%%%%%%%%%%%%%%%%%%%%%%%%%%%%%%%%%%%%%%%%%%%%%%%%%%%%
%%%%%%%%%%%%%%%%%%%%%%%%%%%%%%%%%%%%%%%%%%%%%%%%%%%%%%%%%%%%%%
Summarizing, we have investigated the spin dynamics for one atom
coupled to a high finesse resonator by non-destructively measuring
the atomic state. Continuous probing the
system by this detection method reveals quantum jumps of the atom.
Further, we have measured the normal mode splitting
of the strongly interacting atom-cavity system via the atomic
excitation. Finally, we have observed evidence for an effective atom-atom interaction in spin dynamics of two
atoms simultaneously coupled to the resonator mode. 
Our measurement method fulfills all requirements for a
projective quantum non-demolition measurement of the state of a
single atom \cite{HarocheBook,Braginsky} on a timescale short compared to the inverse jump rate. Operating in a different regime, e.g.~where the effect of atom-field interaction can be deduced from the phase shift of the transmitted probe light \cite{Hood00,Windpassinger2008}, could improve the continuous measurement to the point required for active
feedback onto the quantum state \cite{Feedback}.

%\paragraph{Acknowledgements ---} \approx
We acknowledge financial support by the EC (IP SCALA). SR
acknowledges support from the ``Deutsche Telekom Stiftung" and TK
acknowledges support from the ``Studienstiftung des deutschen
Volkes".
%%%%%%%%%%%%%%%%%%%%%%%%%%%%%%%%%%%%%%%%%%%%%%%%%%%%%%%%%%%%%%


\begin{thebibliography}{20}
\bibitem{footnote}
For the atomic transition frequencies we have taken into account
the AC Stark shift caused by the dipole trap.
\bibitem{footnote1}
For these parameters our system is in the weak excitation regime,
i.e.~the atomic excitation rate is much smaller than the atomic
decay rate.
\end{thebibliography}
\end{document}